\documentclass[aps,prl,twocolumn,showpacs,superscriptaddress,showkeys]{revtex4-1}  
\usepackage{graphicx}  
\usepackage{dcolumn}   
\usepackage{bm}        
\usepackage{amssymb}   
 \usepackage{color}
 \usepackage{amsmath,amsfonts,amssymb}
\usepackage{geometry}

\def\cD{\mathcal{D}}
\def\cH{\mathcal{H}}
\def\cS{\mathcal{S}}
 \def\ep{\varepsilon}

\newtheorem{theorem}{Theorem}

\renewcommand{\leq}{\leqslant}
\renewcommand{\geq}{\geqslant}

\DeclareMathOperator{\vol}{\mathrm{vol}}
\DeclareMathOperator{\tr}{\mathrm{tr}}

\newcommand{\R}{\mathbf{R}}
\newcommand{\C}{\mathbf{C}}

\renewcommand{\P}{\mathbf{P}}

\newcommand{\st}{\  : \ }

\begin{document}

\widetext

\title{Phase transitions for random states and a semicircle law for the partial transpose}

\author{Guillaume Aubrun}
\email{aubrun@math.univ-lyon1.fr}
\affiliation{Institut
Camille Jordan, Universit\'{e} Claude Bernard Lyon 1, 
69622 Villeurbanne CEDEX, France.}

\author{Stanis\l aw J. Szarek}
\email{szarek@cwru.edu}
\affiliation{Case Western Reserve University,
Cleveland, Ohio 44106-7058, USA.}
\affiliation{Institut de Math\'ematiques de Jussieu, Universit\'{e}  Pierre et Marie Curie, 75005 Paris, France}

\author{Deping Ye}
\email{deping.ye@mun.ca}
\affiliation{Department of Mathematics and Statistics, Memorial University of Newfoundland, St. John's, NL, Canada A1C 5S7}


\begin{abstract}
For a system of $N$ identical particles in a
random pure state, there is a threshold $k_0 = k_0(N)\sim N/5$ such that
 two subsystems of $k$ particles each typically share entanglement if 
$k>k_0$,  and  typically do not share entanglement if $k<k_0$. 
By ``random'' we mean here ``uniformly distributed on the sphere of the corresponding Hilbert
space.'' The analogous phase transition for the positive partial transpose (PPT) 
property can be described even more precisely. For example, for $N$ qubits   
the two subsystems of size $k$ are typically in a  PPT state if $k < k_1:=N/4-1/2$ 
and typically in a non-PPT state if $k > k_1$.  Since, for a given state of  the entire system, 
the induced state of a subsystem is given by the partial trace, the above facts 
can be rephrased as properties of random  induced states. 
An important step in the analysis depends on identifying 
the asymptotic spectral density of the partial transposes of such random induced states, 
a result which is interesting in its own right.

\end{abstract}

\maketitle

\section{Introduction}

\vskip-3mm

If all that we know about a quantum system 
is its dimension $n$ (the number of levels) 
and that it is  well isolated from the environment, 
a reasonable model -- or at least a reasonable first guess -- 
for the state of the system is a unit vector selected at random 
from the sphere of an $n$-dimensional complex Hilbert space $\cH$.  
If the system interacts with some part of the environment, represented by 
an ancilla space $\cH_a$, the quantum formalism suggests as a model 
the so-called (random) {\em induced state}, obtained  after partial tracing, 
over $\cH_a$, a random pure state on the space $\cH \otimes \cH_a$. 
The same description  applies if we are primarily interested in a {\em subsystem} 
of an isolated system, the setup that is addressed in the abstract.

The above is just one example of how a random paradigm arises naturally 
in the quantum context. 
In the last few years probabilistic considerations have become a very fruitful approach in
quantum information theory, the highlights being the fundamental paper 
\cite{Hayden2006} by Hayden, Leung and Winter  and, more recently,  Hastings's
proof that suitably chosen random channels provide a counterexample to the additivity
conjecture for classical capacity
of quantum channels \cite{Hasting2009}.

Although random states have been considered for many years, 
their properties (e.g.,  {\em are they typically entangled?}) 
remained elusive. 
In this note we 
describe a reasonably general way to handle such questions.
Of course, the induced state $\rho$ being random, we can not
expect to be able to tell what $\rho$ is. However, we may be
able to infer some properties of $\rho$ if they are {\em generic}
(that is, occur with probability close to $1$). 
 As it turns out, for many natural properties  
a phenomenon of phase transition takes place 
(at least when $\dim \cH$ is sufficiently large):  the generic behavior 
of $\rho$ ``flips''  to the opposite one when $s:=\dim \cH_a$ changes from 
being  a little smaller than certain  threshold dimension $s_0$ 
to being larger than $s_0$.

For simplicity, we will  focus on the  
{\em random induced states} mentioned at the beginning of the Introduction.  
This leads (see \cite{Zycz2,Ben1}) to a natural family of probability measures on $\cD(\cH)$, the
set of states on $\cH$, where $s$, 
the dimension of the ancilla space,  is a parameter.
For specificity, consider $\cH=\C^d \otimes \C^d$ 
and let us concentrate on two properties: {\em entanglement} and {\em positive 
partial transpose} (PPT).   This choice is based, first,  on the importance 
of these concepts and, second, on the differences in their 
respective mathematical  features, 
which allow to present 
the diverse techniques  
needed to handle   
the problems. 

Concerning the importance aspect, we note that detecting and exploiting entanglement 
-- originally discovered in the 1930's \cite{EPR1} --   is a 
central problem in quantum information and quantum computation at least 
since Shor's work \cite{Sh1} on integer factoring. 
Next, the positive partial transpose is the simplest test for entanglement 
(Peres--Horodecki PPT criterion, see \cite{Ho1, Pe1})  
and is at the center of the important {\em distillability conjecture}  \cite{Ho3}, 
a positive answer to which would give a physical/operational meaning 
to the PPT property.  
On the other hand,  
from the  computational complexity 
point of view, verifying the PPT property is easy (just check whether   
the partial transpose of the state $\rho$ is positive semi-definite),   
while deciding whether $\rho$ is entangled is a computationally intractable  (NP-hard) 
problem \cite{Gurvits-NPHard}.

In the special case when $n:=\dim \cH$  equals $s= \dim \cH_a$,
 partial tracing over the ancilla space $\cH_a$ leads to the uniform 
distribution on  $\cD(\C^n)$ 
(i.e.,  uniform with respect to the Lebesgue measure, or Hilbert--Schmidt volume, denoted by $\vol$). 
More generally, when $s \geq n$, the corresponding probability measure $\mu_{n,s}$ 
has a density with respect to the Lebesgue measure on $\cD(\C^n)$ which has
a simple form \cite{Zycz2}
\begin{equation} \label{eq:formula-density} \frac{d\mu_{n,s}}{d\! \vol}(\rho)
= \frac{1}{Z_{n,s}} (\det \rho)^{s-n} ,\end{equation}
where $Z_{n,s}$ is a normalization factor. Note that  \eqref{eq:formula-density}
defines the measure $\mu_{n,s}$ (in particular) for
every real $s \geq n$, 
while the partial trace construction makes sense only for integer values of $s$. 
If $s< n$, the measure $\mu_{n,s}$ is concentrated on the boundary of $\cD(\C^n)$, 
but still can be described analytically.  
Another way to implement these measures is to start from the {\em complex} 
Wishart--Laguerre matrices $W_{n,s}$ ($n\times n$, with $s$ degrees of freedom)  \cite{wishart},
a classical ensemble in statistics and mathematical physics, 
and to normalize them to have trace $1$.  

In spite of the explicitness of the formula \eqref{eq:formula-density}, 
it is not easy to find -- even approximately, and even for $s=n=d^2$ -- the 
probability that a random induced state has PPT or is entangled. 
This is because these traits are not encoded in a simple way  
in the spectral properties of $\rho$.  
It was shown  in \cite{AS1} -- via methods of high-dimensional probability -- that
the proportion of states 
(measured in the sense of $\mu_{n,n}$, i.e., the Legesgue measure) 
that are un-entangled, or {\em separable},  is extremely small in large dimensions. 
This was extended to the case when $s=\dim \cH_a$ is slightly larger than 
$n=d^2$ in \cite{Deping2010},  while, on the other hand, it was proved 
in \cite{Hayden2006} that random 
induced states on $\C^d \otimes \C^d$ are typically separable when 
$s$ is proportional to $n^2=d^4$. The paper  \cite{AS1}  also established 
that un-entangled states are extremely rare even among PPT states 
(again, when $s=n=d^2$). 
However, even such simple question as  ``{\em Does the proportion 
of the PPT states among all states go to $0$ as the dimension increases?},''
originally asked in \cite{Zyczkowski1998}, 
has not been rigorously addressed prior to the work that we describe in this note. 
 The results we summarize 
 go a long way in filling  
the gaps in understanding of the phenomena in question 
(see \cite{Aubrun2011,ASY} for details and references). 
We  show that the threshold 
between entanglement and separability occurs when $s$ is roughly of order 
$n^{3/2}=d^3$, and that  the threshold 
between NPT (i.e., non-PPT) and PPT is when $s \sim 4n = 4d^2$. 

The heuristics behind the
consequences stated in the abstract is now as follows. If we have a system of
$N$ particles (with $D$ levels each) which is in a random pure state,
and two subsystems of $k$ particles each, then the ``joint state''  of the
subsystems is modeled by a random induced  state on $\C^d \otimes \C^d$
with $d=D^k$  and $s = D^{N-2k}$. In particular, the relation $k=N/5$, or
$N=5k$, corresponds exactly to $s=d^3$. A similar argument 
applies to the PPT property. 

Another consequence of the results is that, 
 for a large range of parameters, when the ancilla dimension  $s$ 
is roughly between $4d^2$ and $d^3$, 
a generic random state is both PPT and entangled. 
Such states are {\em bound entangled}, or {\em undistillable} \cite{Ho3}
and, in spite of being entangled, are 
useless for purposes such as teleportation or superdense coding 
(cf. \cite{SST}). 
However, since for 
small systems ($2 \otimes 2$ and $2 \otimes 3$, see \cite{St1, Wo1}) 
PPT is equivalent to separability,
one is tempted to think that bound entangled states are an anomaly, and that 
the PPT property remains a good proxy for separability in higher 
dimensions. 
Our results imply that  this heuristic becomes 
misleading for large systems 
and that 
 PPT and separability are quantitatively very different properties.

While, as we postulated, random induced states form the
most natural family of probability measures on $\cD(\C^n)$, 
our methods are fairly robust and allow handling of
other random models. For example, another popular way
to construct random states is to consider mixtures of pure states.
Our analysis applies to this model as well: 
 if $\nu_{n,s}$ is the distribution of 
$ \frac{1}{s} \sum_{i=1}^s | \psi_i \rangle \langle \psi_i |$,  
where $(\psi_i)$ are independent uniform pure states, then all the results
presented for the measures $\mu_{n,s}$ remain valid \emph{mutatis mutandis}
for  $\nu_{n,s}$.

\section{The results}

\vskip-1mm
We recapitulate the setting:   $n = \dim \cH$,  $\psi$ is a (random) unit 
vector uniformly distributed on the sphere of $\cH \otimes \C^s$, 
and $\rho = \tr_{\C^s} |\psi \rangle \langle \psi|$ is a random state 
on $ \cH$ whose distribution is denoted by $\mu_{n,s}$.  
Further, we assume that $n=d^2 > 1$ and $\cH = \C^d \otimes \C^d$;   
states on $\cH$ will be considered entangled, PPT etc., with respect 
to this particular splitting.  
For definiteness, the partial transpose $\Gamma$ will be the transposition in 
the second factor, i.e., defined (by its action on product states) via  
$(\tau_1\otimes \tau_2)^\Gamma =\tau_1\otimes \tau_2^T$. 

\smallskip The first result 
describes the phase transition between generic entanglement and 
generic separability.  

\begin{theorem} \textnormal{\cite{ASY}}
\label{threshold:separability} There exist effectively computable 
constants $C,c>0$
and a threshold function $s_0=s_0(d)$ satisfying

\vskip-3mm
\begin{equation} \label{eq:bounds-s0} cd^{3} \leq s_0(d) \leq Cd^{3} \log^2 d ,  \end{equation}

\vskip-2mm 
\noindent 
such that if $\rho$ is a random state
on $\C^d \otimes \C^d$ distributed according to the measure
$\mu_{d^2,s}$ and if $\ep > 0$, then

\smallskip 
{\rm (i)} for $s \leq (1-\ep)s_0(d)$ we have 

\smallskip 
\hskip1cm $ \P(\rho \textnormal{ is separable}) \leq 2  \exp(-c(\ep)d^3) $  and 

\smallskip 
{\rm (ii)} for $s \geq (1+\ep)s_0(d)$ we have 

\smallskip
\hskip1cm $\P(\rho \textnormal{ is entangled}) \leq 2 \exp(-c(\ep)s ),$

\smallskip  \noindent
where $c(\ep) >0$  depends only on $\ep$.
\end{theorem}

Let us mention that our methods extend also to the multipartite setting and to ``unbalanced''
systems such as $\C^{d_1} \otimes \C^{d_2} $, ${d_1} \neq {d_2}$ -- see \cite{ASY} for 
precise statements.

The idea behind the proof of Theorem  \ref{threshold:separability}, 
based on  tools from high-dimensional convexity, is quite general 
and can be used to estimate thresholds for other properties of random induced states
(beyond separability), provided the set of states with given property is a convex subset
$\mathcal{K} \subset \cD(\cH)$ and has some minimal invariance properties.
However, in the special case of PPT we have a more precise result. 

\begin{theorem} \textnormal{\cite{Aubrun2011}}
\label{threshold:PPT} 
Let   $\rho$ be a random state on 
$\C^d \otimes \C^d$ distributed according to 
$\mu_{d^2,s}$.  Set $s_1(d)=4d^2$ and let $\ep >0$. 
Then 

\smallskip 
{\rm (i)} for $s \leq (1-\ep)s_1(d)$ we have 

\smallskip 
\hskip1.5cm $ \P(\rho \textnormal{ is PPT}) \leq 2 \exp(-c(\ep)d^2) $  and 

\smallskip 
{{\rm (ii)} for $s \geq (1+\ep)s_1(d)$ we have }

\smallskip
\hskip1.5cm {$\P(\rho \textnormal{ is non-PPT}) \leq 2 \exp(-c(\ep)s),$}

\smallskip  \noindent
where $c(\ep) >0$  depends only on $\ep$.
\end{theorem}

As noted in \cite{Aubrun2011}, it is likely that the sharp estimate in (i) is of order 
$\exp{(-c(\ep) d^4)}$; this conjecture leads to interesting large deviation 
problems for matrices $\rho^\Gamma$. 

\smallskip
The  proof of Theorem \ref{threshold:PPT} (except for the exponential estimates on 
the probabilities, which require a unified approach common to both Theorems) 
depends on methods of random matrix theory  
and, specifically, on the following result that identifies {\em asymptotic 
spectral density} of the partial transpose of random induced states, and 
which is of independent interest. 

\smallskip
If $A$ is a 
Hermitian matrix, we will denote by $\lambda_{\rm max} (A)$ and $\lambda_{\rm min} (A)$ the
largest and the smallest eigenvalues of $A$. If $a \in \R$ and $\sigma>0$, the {\em semicircular
distribution} $\mu_{{\rm SC}(a,\sigma^2)}$ is the probability measure with support $[a-
2\sigma,a + 2\sigma]$ and density   $(2\pi\sigma^2)^{-1} \sqrt{4\sigma^2 -(x-a)^2}$.  We then have

\begin{theorem} \textnormal{\cite{Aubrun2011}}
\label{SC:PPT} Given  $\alpha >0$, let $\rho_d$ be a random mixed state on 
$\C^d \otimes \C^d$ distributed according to 
$\mu_{d^2,\lfloor \alpha d^2 \rfloor}$. Then, as $d$ tends to $+\infty$, the eigenvalue
distributions of $\rho_d^\Gamma$ approaches the deterministic measure 
$\mu_{{\rm SC}(1,1/\alpha)}$ in the following sense: for any interval $I \subset \R$, the proportion
of eigenvalues of $\rho_d^\Gamma$ inside the rescaled interval $\frac{1}{d^2}I$ converges (in probability)
towards $\mu_{{\rm SC}(1,1/\alpha)}(I)$.

Moreover, we also have convergence of the extreme eigenvalues 
$\lambda_{\rm max} (d^2\rho_d^\Gamma)$  and  $\lambda_{\rm min} (d^2\rho_d^\Gamma)$ 
to respectively $1+2/\sqrt{\alpha}$ and $1-2/\sqrt{\alpha}$, 
the endpoints of the support of $\mu_{{\rm SC}(1,1/\alpha)}$.
\end{theorem}

\begin{figure}[htbb]
\includegraphics[width=0.49\columnwidth, angle=-90]{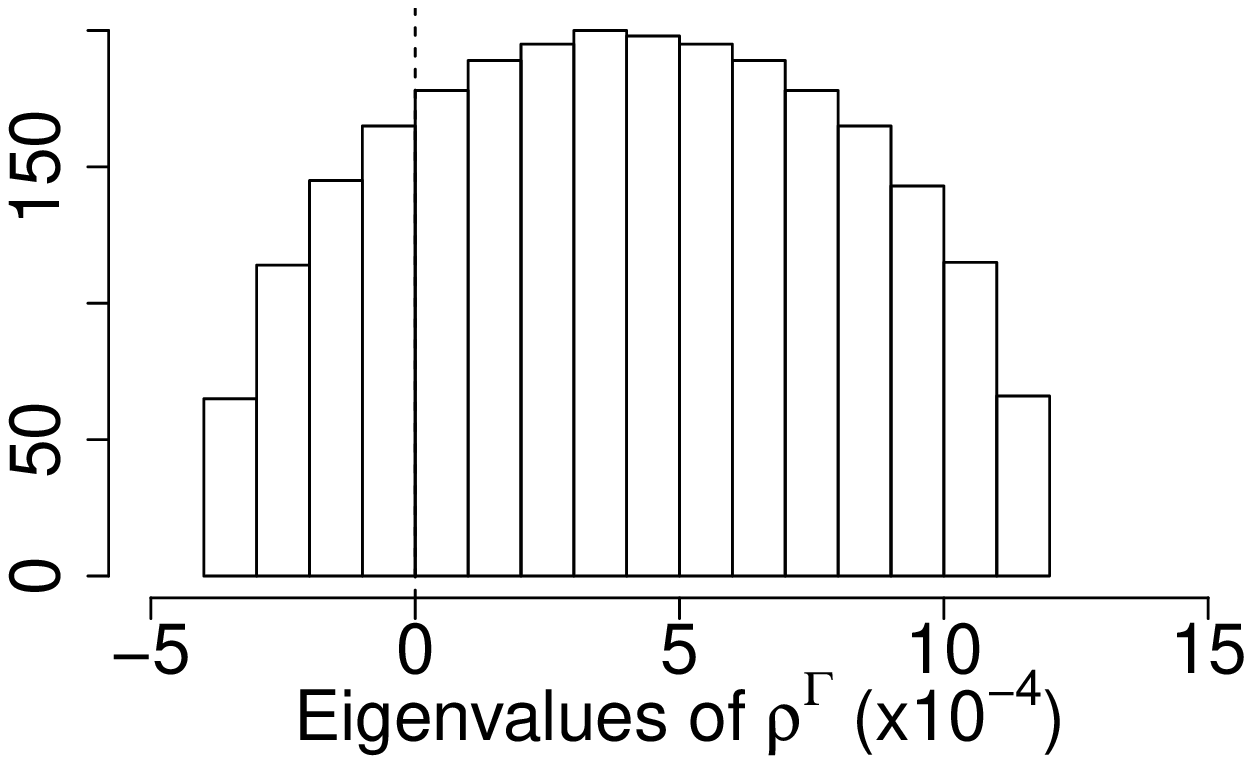}
\vskip6mm
\includegraphics[width=0.49\columnwidth, angle=-90]{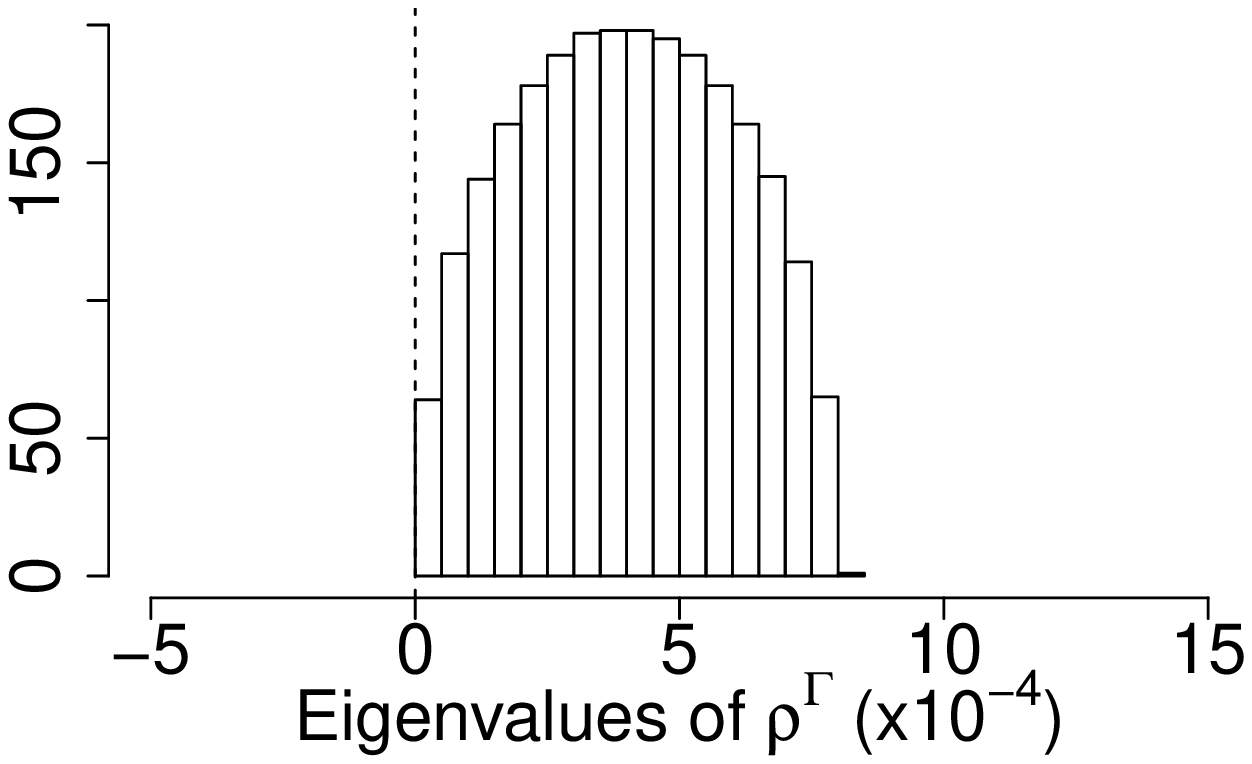}
\vskip3mm
 \caption{Histogram showing distribution of the eigenvalues of $\rho^\Gamma$, 
 where $\rho$ is a random state on   $\C^{50}  \otimes \C^{50}$ 
 chosen according to the distribution $\mu_{2500,2500}$ ($\alpha=1$, top) or
 $\mu_{2500,10000}$ ($\alpha=4$, bottom). 
 In both cases the median eigenvalue is about $\frac{1}{2500}=4\times10^{-4}$.}
 \label{fig}
 \end{figure}

It is easy to numerically ``check''  the conclusion of Theorem \ref{SC:PPT} (this was first noticed in \cite{znidaric}).
For example,  Figure \ref{fig} shows sample   
distributions of eigenvalues of a partially transposed random state on $\C^{50} \otimes \C^{50}$, when $\alpha=1$
and $\alpha=4$ (sample size $1$ in each case).

Because of the link between random induced states and the Wishart ensemble $W_{n,s}$, 
Theorem \ref {SC:PPT}  holds also for that ensemble  (real or complex, 
although it is the complex setting that is most relevant to the quantum theory); 
in that case the rescaling factor $d^2$ is not needed.   
We emphasize that this is rather unexpected since the asymptotic 
spectral density of the Wishart ensemble {\em itself} is given by the Marchenko--Pastur 
distribution  \cite{Marchenko-Pastur 1967}.

\section{The mathematics behind the results}

\vskip-1mm
Although Theorems  \ref{threshold:separability} and  \ref{threshold:PPT} have 
similar statements, the tools used in 
their proofs are very different, which parallels the differences in the 
computational complexity of  PPT  vs. that of entanglement. 
However, combining all the tools is often necessary 
to obtain the full strength of the results. 

We first describe the proof of Theorem \ref{threshold:separability}, which is of geometric nature
and where the concept of \emph{mean width} plays a central role. To present it, let us introduce basic
concepts 
associated to a convex body $K \subset \R^m$ containing the origin in the 
interior (see \cite{Pisier1989} for more background). 
The {\em gauge} of $K$ is the function $\|\cdot\|_K$ defined for
$x \in \R^m$ by  \vskip-5mm
\[ \|x\|_K = \inf \{ t \geq 0 \st x \in tK \} .\]

\vskip-1mm The {\em polar} (or dual) body of $K$ is  defined as
\vskip-5mm \[ K^\circ = \{ y \in \R^m \st \langle x,y \rangle \leq 1 \ \ \forall x \in K \} .\]

\vskip-2mm If $u$ is a vector from the unit sphere $S^{m-1}$, the {\em support
function}  of $K$ in the direction $u$ is $h_K(u):=\max _{x\in K}
\langle x, u\rangle = \|u\|_{K^\circ}$. 
Note that $h_K(u) + h_K(-u)$ is the
distance between the two hyperplanes tangent to $K$ and normal to $u$. 
The mean width of $K$ is~then~defined~as
$$
w (K):=\int _{S^{m-1}}
h_K(u)\,d\sigma(u)=\int _{S^{m-1}} \|u\|_{K^\circ} d\sigma(u), 
$$
where $\,d\sigma$ is the
normalized spherical measure on  $S^{m-1}$. 

In our setting, the relevant convex body is   $K = \cS^\circ$, where $ \cS$ is the set of 
mixed separable states on $\cH=\C^d \otimes \C^d$. 
The ambient space $\R^m$ is the space of self-adjoint trace $1$ operators on $\cH$ (hence
$m=d^4-1$), where the maximally mixed state plays the role of the origin. 
Since  $K^\circ = (\cS^\circ)^\circ= \cS$ (the bipolar theorem) and since 
separability of $\rho$ is equivalent to $\|\rho\|_\cS \leq 1$, the crucial question 
is whether $w(\cS^\circ)$ is smaller or larger than $1$. An analysis of this 
question leads to  the following value of the threshold function 
appearing in Theorem \ref{threshold:separability} 
\vskip-3mm
$$s_0(d) = w(\cS^\circ)^2.
$$
Assertions (i) and (ii) can then be derived from 
concentration of measure, and the
heart of the proof is showing \eqref{eq:bounds-s0}, especially the upper bound.

Determining 
the threshold $s_0(d)$ requires finding the typical value of the gauge 
associated to $\cS$,  computing which -- as we
mentioned -- is a  hard problem. 
We take an indirect route and find the order of magnitude
of the threshold using the machinery of high-dimensional geometry, especially 
the so-called $MM^*$-estimate.

The $MM^*$-estimate (see \cite{FTJ,Pisier1989}) is a general theorem which 
relates the mean width of a convex body and the mean width of its polar.
While the abstract formulation may require an affine change of coordinates, 
in the present situation, because of the symmetries
of $\cS$ (invariance under local unitary conjugations), 
we can deduce via simple representation theory the inequalities
\[ 1 \leq w(\cS) w(\cS^\circ) \leq C \log d ,\]
\vskip-1.5mm
\noindent where $C>0$ is a universal constant. 
Since $w(\cS)$ can be estimated by standard techniques of high-dimensional probability
\cite{AS1}, this allows to establish the order of magnitude of $w(\cS^\circ)$ 
(and hence of $s_0(d)$) up to polylog factors.

As  indicated earlier, the same scheme  yields estimates 
for  the thresholds corresponding to other properties including the PPT, 
but it {\em does not}  allow to recover the 
precise order $4d^2$ appearing in Theorem \ref{threshold:PPT}. 
However, the latter result (except for quantitative estimates on the probabilities, 
which require further work, again based on the concentration of measure) follows readily from Theorem \ref{SC:PPT}, 
which describes very precisely the spectrum of the partial transpose  
of a random induced state: the PPT condition is equivalent to 
$\lambda_{\rm min} (\rho_d^\Gamma)\geq 0$, which is generic if $1-2/\sqrt{\alpha}>0$; 
similarly, $\lambda_{\rm min} (\rho_d^\Gamma)< 0$ is generic 
if $1-2/\sqrt{\alpha}<0$ -- hence $\alpha = \frac{s}{d^2}=4$ is the critical value. 
In turn, to show Theorem \ref{SC:PPT} we use the moment method, a standard technique 
from random matrix theory. 
The idea is to identify the asymptotic spectral density by computing its moments.
This leads to problems in asymptotic combinatorics: 
the moments of semicircular distributions 
are given by the Catalan numbers, corresponding to the dominant 
combinatorial terms, while the statements
about convergence of extreme eigenvalues are proved by refining the calculations 
and carefully estimating   contributions  of lower order combinatorial terms.

\section{Conclusions}

\vskip-1mm 
We  established  that random induced states on $\cH = \C^d \otimes \C^d$ exhibit a phase 
transition phenomenon with respect to the dimension $s$ of the ancilla space. 
We exemplified the phenomenon on two properties:  positive partial transpose, for which the 
threshold value of $s$ is $4d^2$, and entanglement, for which the threshold is $d^3$ 
(up to a polylog factor). This allows to determine whether two subsystems of an 
isolated system typically share (or typically do not share) entanglement when knowing  
only the sizes of those subsystems, and similarly for the PPT property.  
In fact, we provide a ``black box''  approach which applies to many natural properties of quantum states.   
Our results motivate further study of the geometry 
of  sets of  quantum states, and that of large deviation behavior 
of some random matrix ensembles related to quantum information theory.  

We expect  the probabilistic methods to continue to play a major role in quantum theory. 
Indeed, the latter field usually involves high-dimensional objetcs; for example, 
the quantum analogue of a byte 
(a state on $(\C^2)^{\otimes 8}$ -- a qubyte, one may say) 
``lives'' in a space of dimension $2^{16}-1=65535$. While  
this makes numerical schemes  mostly impractical 
(the well-known \emph{curse of dimensionality}), randomness is boosted by  
 the presence of many free parameters 
 (one may call this phenomenon \emph{the blessing of dimensionality}). 
The current level of understanding of these aspects of the theory is arguably 
comparable to that of combinatorics in  the 1950's, when  the power of the probabilistic 
method \cite{ErdosRenyi} began to be appreciated and, subsequently, 
the study of random graphs became an intensive area of research. 

\small 
\medskip \noindent {\em Acknowledgements}: 
 Part of this research was performed 
 during the fall of 2010 while SJS and DY 
 visited the Fields Institute 
and while  GA and SJS visited Institut Mittag-Leffler. 
The research of GA is supported in part by the 
{\itshape Agence Nationale de la Recherche} grants ANR-08-BLAN-0311-03 and ANR 2011-BS01-008-02. 
The research of SJS is supported in part by grants from the 
 {\itshape National Science Foundation (U.S.A.)}, from the  {\itshape U.S.-Israel 
Binational Science Foundation}, and by the second ANR grant listed under GA. 
The research of DY has been initiated with support from the Fields
Institute, the NSERC Discovery Accelerator Supplement
Grant \#315830 from Carleton University,  ERA and
NSERC discovery grants from University of Ottawa, and completed while supported
by a start-up grant from the Memorial University of Newfoundland.

\end{document}